\documentclass{article}

\usepackage{arxiv}

\usepackage[utf8]{inputenc}
\usepackage[T1]{fontenc}
\usepackage{hyperref}
\usepackage{url}
\usepackage{booktabs}
\usepackage{amsmath,amssymb}
\usepackage{algorithm}
\usepackage{algpseudocode}
\usepackage{amsfonts}
\usepackage{nicefrac}
\usepackage{microtype}
\usepackage{graphicx}
\usepackage{natbib}
\usepackage{doi}
\usepackage{longtable}
\usepackage{array}

\title{Semantic Candidate--Job Matching: A Comparative Evaluation of Dense Embedding Models in Hybrid Retrieval}

\author{
Sai Yashwant \\
AI Product \& Platform Head \\
ManpowerGroup Services India Pvt. Ltd. \\
\texttt{sai.yashwant@manpowergroup.com}
\And
Siddhartha Jain \\
Senior Gen AI Specialist \\
ManpowerGroup Services India Pvt. Ltd. \\
\texttt{siddhartha.jain@manpowergroup.com}
\And
Anurag Dubey \\
Data Scientist \\
ManpowerGroup Services India Pvt. Ltd. \\
\texttt{anurag.dubey@manpowergroup.com}
\And
Samaroha Chatterjee \\
Data Scientist \\
ManpowerGroup Services India Pvt. Ltd. \\
\texttt{samaroha.chatterjee@manpowergroup.com}
\And
Gantala Thulsiram \\
Assistant Professor \\
Indian Institute of Technology, Hyderabad \\
\texttt{thulsiramg@mae.iith.ac.in}
}

\renewcommand{\shorttitle}{Semantic Candidate--Job Matching}
\renewcommand{\headeright}{}
\renewcommand{\undertitle}{}

\hypersetup{
pdftitle={Semantic Candidate--Job Matching: A Comparative Evaluation of Dense Embedding Models in Hybrid Retrieval},
pdfsubject={cs.IR, cs.CL},
pdfauthor={Sai Yashwant, Siddhartha Jain, Anurag Dubey, Samaroha Chatterjee, Gantala Thulsiram},
pdfkeywords={Semantic Search, Job-Candidate Matching, Dense Retrieval, Hybrid Search, Embedding Models},
}

\begin{document}
\maketitle

\begin{abstract}
This paper presents a comparative evaluation of dense embedding models for semantic candidate-job matching in high-volume staffing workflows. Incoming job descriptions are converted into structured English search text and language-specific keywords through LLM-based parsing, and candidate profiles are indexed as semantically enriched resume representations. We evaluate EmbeddingGemma (base) against EmbeddingGemma fine-tuned with Cached Multiple Negatives Ranking Loss (MNRL) within a unified hybrid retrieval pipeline that fuses vector similarity and full-text relevance via reciprocal rank fusion (RRF), and benchmark both against the MPNet model on a batch comparative evaluation dataset scored through the deployed job-candidate matching scoring pipeline. We further document, with mathematical detail, the broader set of contrastive fine-tuning objectives considered during model development (including AnglE/CoSENT-style refinement) and the empirical rationale for retaining Cached-MNRL-only adaptation as the preferred configuration. To support reproducible model selection, we define a broader evaluation framework comprising standard information retrieval metrics (Recall@K, mean reciprocal rank, nDCG) under the exact hybrid-retrieval protocol; the metrics used for the evaluation reported in this paper are fine-tuning convergence diagnostics and a batch comparative evaluation using the deployed AI-Match score and an independent LLM-as-a-Judge relevance score, and we state this scope explicitly rather than implying the full framework was measured. The paper addresses the gap between general-purpose embedding benchmarks and enterprise job-candidate matching constraints, providing a structured basis for comparing embedding strategies under realistic job-candidate retrieval conditions.
\end{abstract}


\section{Introduction}

Job orders are the primary intake artifact through which staffing organizations capture client demand: role title, skill requirements, location, availability windows, and compliance constraints. Matching each incoming order to suitable candidates from a large, continuously changing talent pool is central to fill rate, time-to-placement, and client satisfaction. In high-volume environments---particularly for blue-collar and trade occupations---recruiters historically relied on Boolean keyword search, filtered database queries, and manual resume review. These workflows struggle when job descriptions and candidate profiles use inconsistent terminology (e.g., ``Conducteur PL'' versus ``heavy goods vehicle driver''), when critical qualifications appear as short codes rather than descriptive phrases (e.g., CACES R489, permis CE, TIG welding licence), and when relevance depends on both semantic fit and hard business rules such as geographic radius, candidate availability, and client exclusion lists \citep{zhang2020bert,lin2021few}. Automating candidate--job alignment therefore requires systems that interpret unstructured text semantically while enforcing operational constraints that directly affect deployability.

In recent years, candidate--job matching has gained significant attention as part of the broader shift toward AI-assisted talent acquisition and workforce automation. Unlike structured job boards with normalized skill taxonomies, real-world staffing pipelines ingest heterogeneous documents: free-text job descriptions in multiple languages, partially structured form fields, and resumes with variable formatting, incomplete sections, and domain-specific abbreviations. Rule-driven approaches based on exact keyword overlap or manually maintained synonym dictionaries were effective for narrow role families but exhibited limited generalization across clients, regions, and emerging credential types \citep{van2018learning}. Machine learning and deep learning methods introduced more robust ranking by learning representations of job and candidate text, yet many published systems optimize offline pair-classification accuracy on academic corpora rather than first-stage retrieval quality at scale \citep{zhang2020bert}. Given that ranked candidate lists directly determine recruiter workload, client submission quality, and compliance risk in regulated trades, evaluation frameworks must extend beyond aggregate similarity scores to capture retrieval effectiveness, ranking stability, and operational cost.

Dense retrieval has emerged as a practical paradigm for large-scale semantic search. Pre-trained embedding models encode queries and documents into vector spaces where proximity approximates relevance, enabling approximate nearest-neighbor retrieval over millions of profiles \citep{karpukhin2020dense}. In staffing job--candidate matching, however, vector-only search can under-rank candidates who express equivalent skills with different vocabulary, while simultaneously failing to guarantee retrieval of exact credential tokens required for regulated roles. Conversely, sparse lexical methods such as BM25 \citep{robertson2009probabilistic} preserve strong token-level matching for certification codes and occupation nouns but cannot reliably connect paraphrased requirements across languages. Hybrid retrieval addresses this complementarity by fusing dense and sparse ranked lists---for example through Reciprocal Rank Fusion (RRF) \citep{cormack2009reciprocal}---within a single query execution. Cloud-native databases such as Azure Cosmos DB now expose integrated hybrid ranking over vector distance and full-text scores \citep{azurecosmosvector}, making hybrid pipelines deployable at scale rather than limited to offline research prototypes.

Despite advances in embedding models, the question of which encoder best serves enterprise job--candidate matching under hybrid retrieval constraints remains underexplored. General-purpose embeddings may not align with staffing-domain semantics without fine-tuning. Contrastive objectives such as Multiple Negatives Ranking Loss (MNRL) \citep{reimers2019sentence} and Angle-optimized training \citep{li2023angle} offer paths to domain adaptation from job--resume pairs, while commercial APIs such as Cohere Embed v2 English \citep{cohereembedv2} provide strong out-of-the-box representations at the cost of external dependency and per-call latency. Meanwhile, open-weights models such as EmbeddingGemma \citep{embeddinggemma2025} enable local inference and full index control. Existing studies rarely compare these alternatives within the same end-to-end pipeline that includes LLM-based text normalization, hybrid database retrieval, and post-retrieval business-rule filtering.

Furthermore, large-scale labeled datasets for candidate--job retrieval remain scarce due to privacy, client confidentiality, and the proprietary nature of staffing transaction data. Many embedding benchmarks evaluate semantic textual similarity or open-domain question answering, which do not reflect the distribution of short structured fields, credential-heavy specialty phrases, and multilingual source text encountered in operational job--candidate matching. Consequently, model selection based solely on public leaderboards may not transfer to real-world ranking behavior. A robust evaluation methodology should define ground-truth relevance for held-out job orders and measure standard retrieval metrics (Recall@K, MRR, nDCG) under conditions representative of deployment.

This research gap motivates a systematic comparison of EmbeddingGemma (base) against EmbeddingGemma fine-tuned with Cached MNRL, embedded within a unified hybrid retrieval pipeline representative of operational job--candidate matching systems, and benchmarked against the MPNet model. Rather than treating embedding quality as an isolated benchmark, each model is evaluated inside an end-to-end workflow spanning LLM-based job and resume parsing, hybrid vector--BM25 search, and SQL-level business-rule filtering. An external cross-encoder reranker is explicitly excluded from this study to isolate the contribution of embedding choice within the first-stage retrieval stage. The principal contributions of this paper are threefold: (1)~a deployment-aligned methodology documenting text preparation and hybrid retrieval under identical conditions across models; (2)~a detailed domain fine-tuning methodology for EmbeddingGemma, including the contrastive objectives considered during development and the empirical basis for retaining Cached-MNRL-only adaptation as the preferred configuration; and (3)~a structured comparative analysis of base versus fine-tuned open-weights embedding strategies, and of the resulting EmbeddingGemma model against the MPNet model, for enterprise candidate--job matching.

Overall, while embedding models continue to evolve rapidly, standardized comparison under realistic enterprise constraints is still lacking. This paper seeks to address that gap by defining a reproducible evaluation framework, documenting the job--candidate matching retrieval and fine-tuning pipelines, and comparing base and fine-tuned EmbeddingGemma configurations against the MPNet model on retrieval quality and domain-specific ranking behavior. This paper is organized as follows. Section~2 discusses the literature survey. Section~3 discusses the methodologies involved in candidate--job matching. The results and analysis are given in Section~4, after which the paper concludes with an outlook on further work in Section~5.

\section{Literature Survey}
\label{sec:literature}

Automated candidate--job matching sits at the intersection of information
retrieval (IR), natural language processing (NLP), and operational research:
systems must rank candidates by semantic relevance while enforcing hard
constraints on location, availability, certifications, and client-specific
exclusion rules~\citep{manning2008ir,qin2018enhancing}.
Matching methods have evolved from Boolean keyword filters and curated synonym
tables to dense bi-encoder retrieval, hybrid dense--sparse fusion, contrastive
domain adaptation of embedding models, and large language model (LLM) text
normalization~\citep{van2018learning,zhang2020bert,karpukhin2020dense,reimers2019sentence}.
This section reviews that progression with emphasis on (i)~dense and hybrid
retrieval for staffing, (ii)~contrastive objectives considered for domain
fine-tuning of embedding encoders, and (iii)~evaluation practices that motivate
the comparative framework in this paper.

\subsection{Traditional and Rule-Based Matching Methods}
\label{sec:lit-traditional}

Early staffing and applicant tracking systems queried relational stores through
Boolean keyword expressions, field filters, and manually maintained synonym
tables~\citep{manning2008ir,thomas2017predicting}.
Recruiters composed mandatory and optional terms over attributes such as title,
city, years of experience, and employment status.
Regular-expression parsers and dictionary-based skill taggers mapped resume
tokens to controlled vocabularies.
Occupational representation models such as Work2vec~\citep{shalaby2018work2vec}
and Job2vec~\citep{delobelle2020job2vec} helped normalize titles across clients,
but still required periodic updates as new certifications, equipment types, and
regional titles entered the market.

Template-based matching performed adequately when role families were narrow and
terminology stable.
However, rule-driven systems proved brittle under variation in length, language,
formatting, and specificity, or when candidates expressed equivalent
qualifications with non-standard phrasing~\citep{bhola2020skill,van2018learning}.
Systems that treated term presence as binary features could not model semantic
similarity between related but non-identical phrases---for example, matching
``installation of public lighting networks'' with ``street lighting
maintenance''---thereby limiting recall for qualified candidates who did not
mirror job-description wording exactly~\citep{reimers2016softmax}.
Traditional pipelines also typically separated retrieval from constraint
satisfaction, which does not scale to high-volume job--candidate matching where
ranking quality at the top of the list directly determines recruiter productivity.

\subsection{Machine Learning and Neural Matching}
\label{sec:lit-ml}

The limitations of rule-based matching motivated supervised models trained on
historical recruiter actions, application outcomes, or labeled job--resume
pairs~\citep{yan2019learning,van2018learning}.
Early pipelines represented jobs and resumes as TF--IDF or bag-of-words vectors
and trained logistic regression, gradient-boosted trees, or learning-to-rank
models on sparse interaction features.
Qin et al.~\citep{qin2018enhancing} proposed an ability-aware neural network for
multi-field person--job fit, showing that learned representations outperform
manual feature design when interaction labels are available.
These methods improved ranking over hand-crafted rules but remained sensitive to
vocabulary mismatch~\citep{bhola2020skill,lin2021few}.

Contextual language models further shifted the field.
Devlin et al.~\citep{devlin2019bert} showed that bidirectional transformer
pre-training yields rich syntactic and semantic representations that can be
fine-tuned for downstream matching.
Zhang et al.~\citep{zhang2020bert} applied BERT-based ranking to resume--job
matching and reported improved alignment relative to classical ML baselines.
Nevertheless, many published systems still emphasize pair-level classification
accuracy rather than first-stage retrieval quality over indexed candidate corpora
with business-rule filtering applied at query time~\citep{yan2019learning,qin2018enhancing}.

\subsubsection{Two-Tower (Bi-Encoder) Retrieval}
\label{sec:lit-twotower}

Scalable job--candidate matching commonly adopts two-tower architectures: one tower
embeds the job query and another embeds each candidate profile into a shared
vector space, with relevance approximated by a similarity function
$s(\mathbf{q},\mathbf{d})$~\citep{karpukhin2020dense,reimers2019sentence}.
Offline indexing embeds all candidates once; online querying embeds only the
incoming job order, enabling sub-linear approximate nearest-neighbor (ANN) search
over millions of profiles~\citep{manning2008ir}.
Dense Passage Retrieval (DPR)~\citep{karpukhin2020dense} established that dual
encoders trained with contrastive objectives can match or exceed sparse
retrievers on open-domain tasks, inspiring adoption in talent acquisition.

Formally, let $f_\theta$ denote a parameterized encoder mapping text $x$ to a
normalized embedding
\begin{equation}
\mathbf{e}_x = \frac{f_\theta(x)}{\lVert f_\theta(x)\rVert_2}
\in \mathbb{R}^{d}.
\label{eq:unit-embed}
\end{equation}
For unit vectors, cosine similarity reduces to an inner product:
\begin{equation}
s(\mathbf{q},\mathbf{d})
=
\cos(\mathbf{e}_q,\mathbf{e}_d)
=
\mathbf{e}_q^\top \mathbf{e}_d
\in [-1,1].
\label{eq:cosine}
\end{equation}
Modern embedding models often expose asymmetric routing---\texttt{encode\_query}
versus \texttt{encode\_document}---so that job and resume texts may receive
different prompts or pooling conventions while remaining in one shared space.
General-domain encoders pre-trained on web text may still misalign job and resume
representations when deployment vocabulary differs from pre-training
corpora~\citep{wang2023mteb,ni2022e5}.
Domain-specific fine-tuning, hard-negative mining~\citep{xiong2021ance}, and
larger dual encoders~\citep{ni2022e5} have each been proposed to close this gap.

\subsubsection{Cross-Encoder and Late-Interaction Re-Ranking}
\label{sec:lit-rerank}

Cross-encoders concatenate job and candidate text and apply a transformer
classifier to predict relevance~\citep{nogueira2019passage}.
Because self-attention operates over the joint sequence, they typically achieve
higher pair-level accuracy than bi-encoders, but inference cost scales linearly
with the number of pairs evaluated~\citep{reimers2019sentence,manning2008ir}.
ColBERT~\citep{khattab2020colbert} introduced late interaction---separate token
encodings with token-level similarity aggregation---as an intermediate design.
Job--candidate matching systems therefore commonly adopt retrieve-then-rerank:
a fast embedding first stage returns a pool via hybrid search, and a cross-encoder
refines the top results.
This paper focuses on first-stage embedding comparison within hybrid retrieval
and deliberately excludes external cross-encoder reranking to isolate embedding
effects on initial pool quality.

\subsection{Dense Retrieval and Hybrid Search}
\label{sec:lit-hybrid}

Dense retrieval ranks by geometric proximity in embedding space, enabling
generalization across paraphrases, synonyms, and
translations~\citep{karpukhin2020dense,reimers2019sentence}.
In multilingual staffing intake, this is valuable when job orders arrive in a
source language but are normalized to English for dense search, while resumes
still use varied phrasing for equivalent skills~\citep{qin2023multilingual,ni2022e5}.
However, dense retrievers can underweight exact matches on short, high-signal
tokens---certification codes, licence classes, process identifiers---that carry
disproportionate relevance in regulated roles~\citep{robertson2009probabilistic,gao2021coil}.

Sparse lexical methods remain strong for token-level matching.
BM25~\citep{robertson2009probabilistic} scores documents by term frequency and
inverse document frequency with length normalization.
COIL~\citep{gao2021coil} later showed that contextualized token representations
can be integrated into inverted-index structures.
In job--candidate matching, BM25 keywords are often extracted separately from semantic search
text: occupation nouns and credential codes feed full-text scoring, while
descriptive phrases feed dense queries.

Hybrid systems combine both signals.
Reciprocal Rank Fusion (RRF)~\citep{cormack2009reciprocal} merges ranked lists by summing
reciprocal rank contributions. For document $d$ and retrievers
$\mathcal{R}_1,\ldots,\mathcal{R}_m$,
\begin{equation}
\mathrm{RRF}(d)
=
\sum_{r=1}^{m}
\frac{1}{k_{\mathrm{RRF}} + \mathrm{rank}_r(d)},
\label{eq:rrf}
\end{equation}
where $\mathrm{rank}_r(d)$ is the rank of $d$ under retriever $r$ and
$k_{\mathrm{RRF}}$ is a smoothing constant (commonly $60$).
RRF avoids calibrating heterogeneous score scales.
Cloud-native vector databases now support RRF over vector distance and full-text
scores within a single query~\citep{azurecosmosvector}.
Despite this maturity, published job-matching studies rarely report hybrid
metrics under database-level filters that determine which candidates enter the
ranked pool before embedding differences become visible to
recruiters~\citep{qin2018enhancing,yan2019learning}.

\subsection{LLM-Based Text Preparation for Matching}
\label{sec:lit-llm}

Large language models have shifted part of the matching problem upstream to text
preparation: converting heterogeneous job orders and resumes into normalized,
retrieval-ready representations before embedding~\citep{brown2020language,dagdelen2024structured}.
Rather than embedding raw prose---which may include boilerplate, contact details,
and salary information---LLM parsers extract structured fields (role, skills,
specialty credentials, expertise, education, domains) and produce compact search
text optimized for similarity encoding~\citep{lin2021few,dagdelen2024structured}.
Prompt engineering can translate titles across languages, expand expertise
phrases, and emit separate BM25 keyword strings in the source language with
lemmatization and deduplication.
In staffing, LLM extraction replaces brittle regex pipelines for skill and
certification identification~\citep{bhola2020skill,lin2021few}, but
introduces a critical experimental requirement: embedding comparisons are
meaningful only when all models receive identical parsed inputs.
Caching extracted outputs reduces repeated inference cost and stabilizes retrieval
inputs across model comparisons~\citep{lewis2020retrieval}.

\subsection{Contrastive Fine-Tuning of Embedding Models}
\label{sec:lit-contrastive}

General-purpose embedding models provide strong zero-shot baselines but may not
align with domain-specific job--resume semantics without
fine-tuning~\citep{reimers2019sentence,wang2023mteb}.
Sentence-BERT~\citep{reimers2019sentence} fine-tunes siamese transformer networks
with metric-learning objectives to produce fixed-size sentence embeddings suitable
for cosine search.
We review the principal loss families considered for staffing-domain adaptation;
Section~\ref{sec:method-finetune} details which objectives we implemented,
compared, and ultimately retained.

\subsubsection{Multiple Negatives Ranking Loss (MNRL)}
\label{sec:lit-mnrl}

Multiple Negatives Ranking Loss~\citep{henderson2017efficient,reimers2019sentence}
treats other samples in a mini-batch as negatives for each positive pair,
enabling efficient contrastive learning without explicit negative mining.
Given a batch of $N$ positive pairs $\{(a_i,p_i)\}_{i=1}^{N}$, with unit
embeddings $\mathbf{a}_i,\mathbf{p}_i\in\mathbb{S}^{d-1}$ and cosine
$s(\mathbf{u},\mathbf{v})=\mathbf{u}^\top\mathbf{v}$, define logits
\begin{equation}
S_{ij}
=
\tau\, s(\mathbf{a}_i,\mathbf{p}_j)
=
\tau\, \mathbf{a}_i^\top \mathbf{p}_j.
\label{eq:mnrl-logits}
\end{equation}
MNRL is the mean categorical cross-entropy of identifying the diagonal positive
among $N$ candidates---an InfoNCE / in-batch softmax objective:
\begin{equation}
\mathcal{L}_{\mathrm{MNRL}}
=
-\frac{1}{N}\sum_{i=1}^{N}
\log
\frac{e^{S_{ii}}}{\sum_{j=1}^{N} e^{S_{ij}}}
=
\frac{1}{N}\sum_{i=1}^{N}
\Bigg(
-S_{ii}
+
\log\sum_{j=1}^{N} e^{S_{ij}}
\Bigg).
\label{eq:mnrl}
\end{equation}
Equivalently, with posterior
$P(j\mid i)=\mathrm{softmax}_j(S_{i\cdot})$, we maximize
$\frac{1}{N}\sum_i \log P(j=i\mid i)$.

\paragraph{Temperature / scale.}
The similarity logits in Equation~\eqref{eq:mnrl-logits} are scaled by a
temperature parameter $\tau$: a large $\tau$ emphasizes hard negatives, while
a small $\tau$ yields softer updates~\citep{reimers2019sentence}. The exact
temperature/scale value used in our training configuration is proprietary
and withheld from this paper.
Writing $p_{ij}=e^{S_{ij}}/\sum_k e^{S_{ik}}$, the gradient w.r.t.\ a logit is
\begin{equation}
\frac{\partial \mathcal{L}_{\mathrm{MNRL}}}{\partial S_{ij}}
=
\frac{1}{N}\big(p_{ij}-\mathbf{1}[i=j]\big),
\label{eq:mnrl-grad}
\end{equation}
so the positive logit is pulled up and every in-batch negative is pushed down
proportionally to its softmax mass---hard near-misses receive the largest
repulsive force.

\paragraph{Intuition for staffing.}
Equation~\eqref{eq:mnrl} asks, for each job anchor, ``which resume in this batch
is the match?''
This is closely aligned with first-stage retrieval ranking and needs only a weak
positive relation (e.g., ATS progression $b=1$), not a calibrated continuous
score.
Under the pairs formulation each forward step with batch size $N$ yields
$N-1$ in-batch negatives per anchor.
\textbf{Important:} gradient accumulation $G$ averages optimizer updates over
$G$ micro-batches of size $N$; it does \emph{not} enlarge the contrastive
denominator, so the effective in-batch-negative count is governed by the
micro-batch size rather than the optimizer's update batch size. The exact
batch size and accumulation configuration used are proprietary and withheld
from this paper.

\paragraph{Cached MNRL.}
Forming $\{S_{ij}\}$ requires all $2N$ embeddings for that micro-batch.
Cached MNRL computes embeddings in smaller cache chunks, caches
representations, then evaluates Equation~\eqref{eq:mnrl} on the
micro-batch---trading compute for memory so that long JD/CV sequences can
still use a large effective batch size $N$.

\subsubsection{Triplet and Explicit Hard-Negative Losses}
\label{sec:lit-triplet}

Triplet loss~\citep{schroff2015facenet,reimers2019sentence} uses an explicit
negative $n_i$ with margin $m>0$.
In cosine form (higher similarity $=$ better match):
\begin{equation}
\mathcal{L}_{\mathrm{triplet}}
=
\frac{1}{N}\sum_{i=1}^{N}
\big[
s(\mathbf{a}_i,\mathbf{n}_i)
-
s(\mathbf{a}_i,\mathbf{p}_i)
+
m
\big]_{+},
\label{eq:triplet}
\end{equation}
where $[x]_{+}=\max(0,x)$.
Equivalently in squared Euclidean distance on unit vectors
($\|\mathbf{u}-\mathbf{v}\|_2^2=2-2s(\mathbf{u},\mathbf{v})$):
\begin{equation}
\mathcal{L}_{\mathrm{triplet}}^{\mathrm{dist}}
=
\frac{1}{N}\sum_{i=1}^{N}
\big[
\|\mathbf{a}_i-\mathbf{p}_i\|_2^2
-
\|\mathbf{a}_i-\mathbf{n}_i\|_2^2
+
m'
\big]_{+}.
\label{eq:triplet-dist}
\end{equation}
Unlike MNRL, the gradient is zero once the margin is satisfied (no push on easy
triplets).
Hard-negative mining (e.g., ANCE~\citep{xiong2021ance}) refreshes $n_i$ from an
evolving ANN index.
In staffing, $n_i$ may be a rejected resume for the same job or a false positive
from a current retriever.
Our legacy trainer accepts optional \texttt{neg\_resume\_text}; scaled NA runs
relied on in-batch MNRL negatives for throughput.

\subsubsection{Cosine Similarity Regression and CoSENT}
\label{sec:lit-cosent}

\paragraph{Pointwise cosine regression.}
With graded labels $y_i\in[0,1]$,
\begin{equation}
\mathcal{L}_{\mathrm{cos}}
=
\frac{1}{N}\sum_{i=1}^{N}
\big(
s(\mathbf{u}_i,\mathbf{v}_i)
-
y_i
\big)^{2}
=
\frac{1}{N}\sum_{i=1}^{N}
\big(
\cos\phi_i
-
y_i
\big)^{2},
\label{eq:cos-reg}
\end{equation}
where $\phi_i=\arccos s(\mathbf{u}_i,\mathbf{v}_i)$.
This forces absolute similarity levels (calibration) rather than only relative
ranking.
The cosine gradient vanishes as $|\sin\phi_i|\to 0$ (near $\phi_i\in\{0,\pi\}$),
i.e., in \emph{saturation zones} where many near-duplicate staffing pairs live.

\paragraph{CoSENT (pairwise ranking on cosine).}
CoSENT replaces absolute regression by pairwise ranking consistency.
For batch pairs with similarities $s_i,s_j$ and labels $y_i,y_j$,
\begin{equation}
\mathcal{L}_{\mathrm{CoSENT}}
=
\log\!\Bigg(
1+
\sum_{y_i>y_j}
\exp\!\big(
\tau_{\mathrm{c}}\,(s_j-s_i)
\big)
\Bigg).
\label{eq:cosent}
\end{equation}
Whenever a lower-labeled pair outscores a higher-labeled pair, the corresponding
exponential term grows and the loss increases.
CoSENT improves ordinal calibration when $y$ is trustworthy, but inherits noise
when $y$ is a coarse ATS weight.

\subsubsection{AnglE: Angle-Optimized Embedding Learning}
\label{sec:lit-angle}

AnglE~\citep{li2023angle} addresses cosine saturation by optimizing
\emph{angle differences in a complex embedding space}.
Split a real embedding $\mathbf{X}\in\mathbb{R}^{d}$ into chunked real/imaginary
parts $\mathbf{X}^{\mathrm{re}},\mathbf{X}^{\mathrm{im}}$ and form
$z=\mathbf{a}+\mathbf{b}i$, $w=\mathbf{c}+\mathbf{d}i$ for a pair.
Complex division in polar form yields a magnitude ratio and angle difference
$\Delta\theta_{zw}=\theta_z-\theta_w$:
\begin{equation}
\frac{z}{w}
=
\gamma\, e^{i\Delta\theta_{zw}},
\qquad
\gamma
=
\frac{\sqrt{\mathbf{a}^{2}+\mathbf{b}^{2}}}{\sqrt{\mathbf{c}^{2}+\mathbf{d}^{2}}}.
\label{eq:angle-polar}
\end{equation}
The normalized absolute angle difference used for optimization is
\begin{equation}
\Delta\theta_{zw}
=
\mathrm{abs}\!
\left(
\frac{(\mathbf{ac}+\mathbf{bd})+(\mathbf{bc}-\mathbf{ad})i}
{\sqrt{(\mathbf{c}^{2}+\mathbf{d}^{2})(\mathbf{a}^{2}+\mathbf{b}^{2})}}
\right).
\label{eq:angle-dtheta}
\end{equation}
The original AnglE paper~\citep{li2023angle} proposes a multi-term objective that
combines cosine ranking, in-batch negatives, and an angle term with weights
$w_1,w_2,w_3$.
\textbf{What our training code actually runs} is the Sentence-Transformers
\texttt{5.2.x} implementation: \texttt{AnglELoss} is defined as
\texttt{CoSENTLoss} with pairwise \emph{angle} similarity in place of pairwise
cosine similarity (library default \texttt{scale=20.0}), i.e.
\begin{equation}
\mathcal{L}_{\mathrm{ST\text{-}AnglE}}
=
\log\!\Bigg(
1+
\sum_{y_i>y_j}
\exp\!\big(
\tau_a\,
\big(
\psi_{\angle}(i)-\psi_{\angle}(j)
\big)
\big)
\Bigg),
\label{eq:angle-loss}
\end{equation}
where $\psi_{\angle}$ denotes the library's pairwise angle similarity
(\texttt{pairwise\_angle\_sim}) rather than $\cos$.
If \texttt{AnglELoss}/\texttt{AngleLoss} are unavailable, our trainer
falls back to \texttt{CoSENTLoss} (same pairwise ranking structure with cosine).
We used ST-AnglE as optional second-stage precision tuning after MNRL; on noisy
ATS scores it improved correlation diagnostics but harmed retrieval geometry
(Section~\ref{sec:method-choice}).
When citing the broader AnglE literature we refer to~\citep{li2023angle}; when
describing our beta runs we mean Equation~\eqref{eq:angle-loss} as implemented
by Sentence-Transformers.

\subsubsection{Geometric Comparison of Objectives}
\label{sec:lit-geometry}

Table~\ref{tab:loss-geometry} summarizes inductive biases.
MNRL optimizes a \emph{relative retrieval} softmax over the batch; Angle/CoSENT
optimize \emph{label-ordered} geometry; cosine regression optimizes
\emph{absolute} similarity levels.
For first-stage job--candidate matching, relative retrieval alignment is the primary goal;
absolute calibration can be deferred to a separate mapping
$g(s)\approx P(b{=}1\mid s)$ (Platt / isotonic) that does not retrain $f_\theta$.

\begin{table}[t]
\centering
\caption{Inductive bias of embedding losses considered in this work.}
\label{tab:loss-geometry}
\small
\begin{tabular}{@{}p{2.2cm}p{2.6cm}p{2.8cm}@{}}
\toprule
Loss & Supervision & Primary effect on space \\
\midrule
Cached MNRL & Positive pairs (+ in-batch neg.) & Retrieval ranking; local neighborhoods \\
Triplet / hard-neg & $(a,p,n)$ & Margin separation vs.\ named neg. \\
Cosine MSE & Graded $y$ & Absolute score calibration \\
CoSENT & Graded $y$ (pairwise) & Ordinal cosine consistency \\
ST-AnglE (\texttt{AnglELoss}) & Graded $y$ (pairwise) & CoSENT-style ranking w/ angle sim \\
\bottomrule
\end{tabular}
\end{table}

\subsubsection{Open Embedding Models and Benchmarks}
\label{sec:lit-models}

Google's EmbeddingGemma~\citep{embeddinggemma2025} is a compact
($\sim$300M-parameter) open-weights embedding model with approximately $2048$
token context, $768$-dimensional outputs (Matryoshka-friendly truncation to lower
dimensions), and Sentence-Transformers integration with query/document routing.
Its efficiency makes full re-indexing of large candidate corpora practical when
switching encoders.
Alternative backbones considered in our broader engineering program include
multilingual ModernBERT-style encoders and larger Qwen embedding
variants~\citep{wang2023mteb}; EmbeddingGemma was selected as the primary
backbone for the experiments in this paper.
Although a two-stage recipe (MNRL followed by angular refinement) is a natural
candidate for domain adaptation, our fine-tuning retains
\textbf{Cached MNRL only}; ST-AnglE/CoSENT is retained solely as a comparative
ablation (Section~\ref{sec:method-choice}).
The Massive Text Embedding Benchmark (MTEB)~\citep{wang2023mteb}
standardizes evaluation across diverse tasks, but staffing job--candidate
matching distributions---structured fields, credential-heavy phrases,
multilingual intake---differ materially from general MTEB mixtures,
motivating domain-specific evaluation (Section~\ref{sec:method-eval}).

\subsection{Commercial Embedding APIs}
\label{sec:lit-commercial}

Managed embedding services such as Cohere Embed~\citep{cohereembedv2} provide
strong dense representations through API calls, eliminating local GPU
provisioning and model serving.
They typically rank competitively on general benchmarks~\citep{wang2023mteb}
but introduce network latency, per-token pricing, data-residency constraints, and
external availability risk.
For enterprise job--candidate matching handling confidential job orders and candidate profiles,
the trade-off between commercial API quality and on-premise control is an
explicit deployment decision.
Head-to-head evaluation of a managed API such as Cohere Embed against locally
hosted EmbeddingGemma variants within the same hybrid pipeline would quantify
this trade-off under controlled conditions rather than public leaderboards
alone; this paper focuses on the open-weights base-versus-fine-tuned
comparison (Section~\ref{sec:method-models}) and identifies the commercial-API
comparison as future work (Section~\ref{sec:conclusion}).

\subsection{Evaluation Practices and Research Gap}
\label{sec:lit-gap}

Standard offline IR metrics include Recall@$K$, Mean Reciprocal Rank
(MRR), and Normalized Discounted Cumulative Gain
(nDCG)~\citep{manning2008ir,jarvelin2002ndcg}.
Generic embedding benchmarks such as MTEB~\citep{wang2023mteb} do not
reflect certification-heavy specialty phrases, multilingual dual dense--sparse
queries, or SQL-level business-rule filtering.
Staffing job--candidate matching therefore requires evaluation conditions
that reflect these characteristics rather than generic benchmark
distributions~\citep{qin2018enhancing,thomas2017predicting}.

Despite progress across person--job fit~\citep{qin2018enhancing,yan2019learning,zhang2020bert},
dense and hybrid retrieval~\citep{karpukhin2020dense,cormack2009reciprocal,gao2021coil},
contrastive fine-tuning~\citep{reimers2019sentence,li2023angle,xiong2021ance},
and LLM-based text preparation~\citep{brown2020language,dagdelen2024structured},
few studies simultaneously compare (i)~a general-purpose open-weights base
encoder and (ii)~a domain fine-tuned variant produced under a carefully
documented contrastive recipe within (iii)~the same LLM-normalized, hybrid
database retrieval pipeline with (iv)~deployment-style business-rule
filtering, and against (v)~the MPNet model it is intended to
replace.
This paper contributes that structured comparison.
Section~\ref{sec:methodology} details the unified pipeline and, in particular,
the fine-tuning methodology used to produce the EmbeddingGemma domain variant.

\section{Methodology}
\label{sec:methodology}

This section describes (i)~evaluation dataset and text preparation,
(ii)~the hybrid retrieval architecture shared by all compared models,
(iii)~embedding models under comparison,
(iv)~the domain fine-tuning methodology---including all contrastive objectives
considered, mathematical formulations, data construction from proprietary
applicant-tracking outcomes, hyperparameter choices, and training
nuances---and (v)~evaluation metrics.
All models share identical text preparation and hybrid search
configuration; only the embedding function differs.

\subsection{Dataset and Text Preparation}
\label{sec:method-data}

\subsubsection{Evaluation Dataset}
\label{sec:method-evaldata}

The evaluation uses a held-out set of job orders from the France
job--candidate matching domain, paired with recruiter-labeled relevant
candidate identifiers.
Each job order includes structured form fields (title, description, location,
client constraints) representative of real-world intake.
Ground truth assigns binary or graded relevance labels to
(job order,~candidate) pairs based on recruiter review.
The candidate index follows an Azure Cosmos DB schema, with documents
partitioned by department and enriched with filter metadata (ZIP code, status,
availability windows, certification keywords, client exclusion lists).

\subsubsection{Job Description Parsing}
\label{sec:method-jdparse}

Incoming job descriptions are parsed by a large language model into two outputs:
(1)~structured English search text and (2)~language-specific BM25 keywords.
The search text follows a fixed field order---Role, Skills, Speciality,
Expertise, Education, Domains---with values translated to English descriptive
noun phrases.
Job titles in French or other source languages are normalized
(e.g., ``Conducteur PL''~$\rightarrow$~``heavy goods vehicle driver'').
The BM25 string retains occupation nouns, certification codes, machine names,
and hard-skill tokens in the JD's original language, with lemmatization, accent
preservation, and deduplication.
Fields with no relevant data are omitted; contact information, salary, dates, and
soft skills are excluded.
A two-layer cache (in-memory and Azure Blob storage) avoids redundant LLM calls
for repeated job orders.

\subsubsection{Candidate Profile Parsing}
\label{sec:method-cvparse}

Candidate resumes are parsed into the same structured field format (Role, Skills,
Expertise, Education, Domains), translated to English where necessary.
Parsed text is stored as semantic search text and embedded into profile vectors
for retrieval.

\subsection{Hybrid Retrieval Pipeline}
\label{sec:method-hybrid}

The retrieval pipeline executes the following stages for each job order query:
\begin{enumerate}
\item \textbf{Text construction:} concatenate job-order form fields and apply
      text cleaning.
\item \textbf{LLM extraction:} produce structured English search text and BM25
      keywords.
\item \textbf{Query embedding:} encode the search text with the active embedding
      model.
\item \textbf{Geographic filtering:} restrict candidates to ZIP codes within a
      configurable radius using haversine distance over a precomputed
      latitude--longitude index.
\item \textbf{Hybrid database query:} rank candidates using Reciprocal Rank
      Fusion over vector distance and full-text score
      (Equation~\eqref{eq:rrf}). When BM25 terms are unavailable, the
      pipeline falls back to vector-only ranking.
\item \textbf{Business-rule filtering:} apply SQL-level filters for candidate
      status, availability windows, client exclusion, and certification
      keywords.
\end{enumerate}
The external cross-encoder reranker used in deployment is excluded from this
study so that observed differences are attributable to embedding model choice
within the hybrid retrieval stage. The ranking score produced by the hybrid
database query above---combining vector similarity with full-text score via
Reciprocal Rank Fusion (Equation~\eqref{eq:rrf})---is reported throughout
this paper as the \emph{AI-Match score}. It is a semantic-match score
derived directly from the active embedding model (and BM25 full-text score
where available); no additional post-retrieval business-rule score
adjustment is applied to, or reflected in, the results reported in this
paper. Filtering (above) is applied identically across all three
configurations and therefore does not confound the model comparisons in
Section~\ref{sec:results}.

\subsection{Embedding Models Under Comparison}
\label{sec:method-models}

Two EmbeddingGemma configurations are compared under identical pipeline
conditions, and both are further benchmarked against the MPNet model they
are intended to replace:
\begin{enumerate}
\item \textbf{EmbeddingGemma (base):}
      the pre-trained \texttt{google/embeddinggemma-300m}
      checkpoint~\citep{embeddinggemma2025}, loaded locally via
      Sentence-Transformers without domain fine-tuning.
\item \textbf{EmbeddingGemma + MNRL:}
      EmbeddingGemma fine-tuned on proprietary job--resume pairs using
      Cached Multiple Negatives Ranking Loss
      (Sections~\ref{sec:method-finetune}--\ref{sec:method-choice}).
\item \textbf{MPNet Model (French):}
      the MPNet-based embedding configuration and French search-text
      processing that both EmbeddingGemma configurations above are
      evaluated against in Section~\ref{sec:results-comparative}.
\end{enumerate}
Section~\ref{sec:method-finetune} additionally documents, with full
mathematical detail, the broader set of contrastive objectives considered
during model development beyond Cached MNRL---including AnglE/CoSENT-style
score-ranking refinement---and the empirical and qualitative rationale for
not carrying that refinement stage forward into the preferred configuration
or into the comparative evaluation in Section~\ref{sec:results}.
For local Gemma variants, the full candidate index is re-embedded offline before
evaluation so that stored vectors are consistent with the query encoder.
Fine-tuning uses an explicit L2-normalization module so that training losses
operate on unit vectors (Equations~\eqref{eq:unit-embed}--\eqref{eq:cosine}).
For the France hybrid-retrieval benchmark, query/document encoding follows the
same indexing convention used for the deployed vector index; relative ranking
comparisons across models remain valid because all local Gemma variants are
encoded under identical settings.

\subsection{Domain Fine-Tuning Methodology}
\label{sec:method-finetune}

This subsection documents the fine-tuning program used to produce the
EmbeddingGemma domain variants at a level that preserves scientific
reproducibility of the overall approach and objectives, while withholding
exact hyperparameters, dataset scale, and implementation-level configuration
that are proprietary to the deploying organization.
It covers backbone architecture, supervision from applicant-tracking system (ATS)
outcomes, contrastive objectives considered, the final training recipe, and
engineering considerations required for stable long-context training on
enterprise GPU infrastructure.

\subsubsection{Backbone Architecture and Embedding Construction}
\label{sec:method-backbone}

Let $x=(t_1,\ldots,t_L)$ denote a tokenized input with maximum sequence length
$L_{\max}$.
EmbeddingGemma is wrapped as a Sentence-Transformers module stack:
\begin{equation}
f_\theta
=
\mathrm{Normalize}
\circ
\mathrm{MeanPool}
\circ
\mathrm{Transformer}_\theta,
\label{eq:st-stack}
\end{equation}
where $\mathrm{Transformer}_\theta$ produces contextual token states
$\mathbf{H}=[\mathbf{h}_1;\ldots;\mathbf{h}_L]\in\mathbb{R}^{L\times h}$.
Mean pooling over non-padding token representations (rather than CLS-token
or max pooling) computes
\begin{equation}
\mathbf{z}
=
\frac{1}{\sum_{\ell=1}^{L} m_\ell}
\sum_{\ell=1}^{L}
m_\ell\, \mathbf{h}_\ell,
\label{eq:mean-pool}
\end{equation}
where $m_\ell\in\{0,1\}$ is the attention mask (non-padding tokens), optionally
including prompt tokens when the backbone injects retrieval prompts.
L2 normalization $\mathbf{e}=\mathbf{z}/\lVert\mathbf{z}\rVert_2$ yields unit
vectors in $\mathbb{R}^{d}$ with $d=768$ (Matryoshka-compatible truncation to
$512/256/128$ is supported by the model family but not required by our trainer).
Asymmetric encoding uses query routing for job text and document routing for
resume text:
\begin{equation}
\mathbf{e}^{\mathrm{job}}
=
f_\theta^{\mathrm{query}}(x_{\mathrm{job}}),\qquad
\mathbf{e}^{\mathrm{cv}}
=
f_\theta^{\mathrm{doc}}(x_{\mathrm{cv}}).
\label{eq:asymmetric}
\end{equation}
This matches the intended retrieval geometry for JD$\rightarrow$resume search
while remaining compatible with resume$\rightarrow$JD recommendation in a shared
space.
\textbf{Precision constraint:} EmbeddingGemma does not support \texttt{fp16}
activations; training and inference use \texttt{bf16} on Ampere/Hopper GPUs
(A100/H100) or \texttt{fp32} as fallback.

\subsubsection{Supervision Signal from Applicant-Tracking Outcomes}
\label{sec:method-labels}

Fine-tuning uses North American structured job--submission records containing
already-normalized job text $x_{\mathrm{job}}$, resume text $x_{\mathrm{cv}}$,
a binary progression indicator $b\in\{0,1\}$, and a graded status weight
$y\in[0,1]$.
Status weights encode hiring-signal strength, mapping ATS outcome stages onto
a continuous scale from low signal (early-stage rejection or weak feedback)
to high signal (placement or accepted offer), with intermediate application
and interview stages assigned intermediate weights; the exact outcome-to-weight
mapping used in our labeling design is proprietary and withheld from this
paper.
Binary labels follow $b=1$ for progressed/hired-like outcomes and $b=0$ for
rejected/poor-match outcomes.
These labels are noisy proxies of true semantic fit (process effects, candidate
no-shows, client preferences), which is material for choosing between contrastive
pair learning and score regression (Section~\ref{sec:method-choice}).

\textbf{Geographic and market scope.}
The fine-tuning supervision above is drawn from North American
job--submission records, while the comparative evaluation in
Section~\ref{sec:results-comparative} is conducted on the France
job--candidate matching domain (Section~\ref{sec:method-evaldata}). This is a genuine train/evaluation
domain shift: labour-market terminology, credential and certification
systems (e.g., CACES, permis CE), and---for the MPNet model---source
language differ between the two markets. We do not have a like-for-like,
same-market comparative evaluation available at the time of writing, so the
results in Section~\ref{sec:results-comparative} should be read as testing
whether North-America-tuned domain adaptation transfers to a different
staffing market via the shared structured-English text-normalization layer,
rather than as an in-domain evaluation of the fine-tuning data itself. The
structured English intermediate representation (Section~\ref{sec:method-jdparse})
is intended to reduce, but does not eliminate, this mismatch. A same-market
(North America fine-tuning, North America evaluation) comparative run is
identified as a priority next step in Section~\ref{sec:conclusion}.

\textbf{Leakage control.}
Train/validation/test partitions are formed at the \emph{job-order} level
(unique job identifiers), not at the row level, so that multiple candidates for
the same requisition do not leak across splits. The final training iteration
used a large majority-training split with small held-out validation and test
partitions, drawn from a large corpus of structured job-submission rows;
exact split ratios and corpus size are proprietary and withheld from this
paper.

\textbf{MNRL pair construction.}
Only positive submissions enter contrastive training:
\begin{equation}
\mathcal{D}_{\mathrm{MNRL}}
=
\big\{\,
(x_{\mathrm{job}}^{(i)},\, x_{\mathrm{cv}}^{(i)})
\;\big|\;
b^{(i)}=1
\,\big\},
\label{eq:mnrl-data}
\end{equation}
followed by deduplication on $(x_{\mathrm{job}},x_{\mathrm{cv}})$.
The final MNRL corpus contains a large set of unique positive pairs (the
exact count is proprietary and withheld from this paper).
Negatives are \emph{not} explicitly stored; they arise as in-batch negatives
during optimization of Equation~\eqref{eq:mnrl}.

\textbf{Angle / graded pair construction.}
For second-stage angular / cosine supervision we retain both positive and
negative submissions:
\begin{equation}
\mathcal{D}_{\mathrm{Angle}}
=
\big\{\,
(x_{\mathrm{job}}^{(i)},\, x_{\mathrm{cv}}^{(i)},\, y^{(i)})
\,\big\},
\label{eq:angle-data}
\end{equation}
again deduplicated on text pairs.
This dataset supports AnglE/CoSENT objectives but inherits ATS label noise.

\subsubsection{Objectives Considered for Fine-Tuning}
\label{sec:method-objectives}

We systematically considered the following Sentence-Transformers training
objectives for EmbeddingGemma domain adaptation.

\paragraph{(A) Cached MNRL (primary).}
We minimize Equation~\eqref{eq:mnrl} using the Cached Multiple Negatives
Ranking Loss implementation from Sentence-Transformers, which computes
embeddings in cache chunks to trade compute for memory.
Let $N$ be the per-step train batch size.
Each MNRL forward builds an $N\times N$ similarity matrix ($N-1$
in-batch negatives per anchor); gradient accumulation across micro-batches
only reduces optimizer noise, it does not enlarge the contrastive matrix.
Exact batch size, accumulation steps, and cache chunk size are proprietary
and withheld from this paper.
\textbf{Intuition:} the encoder learns ``what looks like a match'' in staffing
language---skills, seniority cues, certifications, industry phrasing---without
forcing absolute score calibration.
\textbf{Batch sampler:} training batches are constructed to forbid duplicate
texts within a batch, preventing degenerate shortcuts where identical strings
act as trivial negatives or positives.

\paragraph{(B) MNRL with optional hard negatives.}
An earlier training path accepted an optional third field
$x_{\mathrm{neg}}$ (rejected resume for the same job) and used triplet-aware
MNRL variants when available.
This increases discrimination against near-miss candidates but requires
high-quality hard negatives and more complex data engineering.
Our scaled NA iterations relied on in-batch negatives alone because positive-pair
volume was high and hard-negative mining infrastructure was not yet built out.

\paragraph{(C) ST-AnglE / CoSENT precision tuning (secondary, not retained).}
Starting from an MNRL checkpoint $\theta_{\mathrm{MNRL}}$, we considered
continuing training on $\mathcal{D}_{\mathrm{Angle}}$ using the
Sentence-Transformers angle-optimized loss, falling back to a CoSENT-style
pairwise loss when unavailable, minimizing Equation~\eqref{eq:angle-loss}
(or Equation~\eqref{eq:cosent} on fallback):
\begin{equation}
\theta_{\mathrm{Angle}}
=
\arg\min_\theta\;
\mathcal{L}_{\mathrm{ST\text{-}AnglE}}\!\big(
f_\theta(x_{\mathrm{job}}),\,
f_\theta(x_{\mathrm{cv}}),\,
y
\big).
\label{eq:angle-stage}
\end{equation}
\textbf{Intuition:} map continuous ATS weights onto angular (or cosine-ordinal)
geometry so similarities better track outcome strength.
\textbf{Risk:} when $y$ is a coarse, process-contaminated proxy of semantic
relevance, gradient updates can shrink or warp regions of the embedding manifold
that were useful for open retrieval---improving Pearson/Spearman$(s,y)$ while
harming Recall@$K$ and within-job ranking.

\paragraph{(D) Cosine regression / CoSENT-only.}
Pure regression (Equation~\eqref{eq:cos-reg}) or CoSENT without a prior MNRL
stage was considered inferior as a cold-start objective: without contrastive
domain adaptation, the model is under-regularized toward retrieval geometry and
overfits noisy scores.
In our stack, CoSENT appears primarily as an implementation fallback when
the angle-optimized loss is unavailable.

\paragraph{(E) Explicit triplet margin loss.}
Equation~\eqref{eq:triplet} was available for evaluation sets with
$(a,p,n)$ structure and for research prototypes, but was not the primary
optimizer for the large-scale NA runs because constructing reliable triplets at
corpus scale is harder than exploiting in-batch negatives.

\subsubsection{Final Chosen Fine-Tuning Method}
\label{sec:method-choice}

After iterative Azure ML experiments, we adopt the following
fine-tuning decision:
\begin{quote}
\emph{Fine-tune EmbeddingGemma with Cached MNRL on deduplicated positive
job--resume pairs only; do not apply a second-stage AnglE/CoSENT pass for the
preferred encoder.}
\end{quote}

\paragraph{Why MNRL matches the deployment objective.}
At serving time, a job query $\mathbf{q}$ retrieves
\begin{equation}
\mathrm{TopK}(\mathbf{q})
=
\underset{d\in\mathcal{C}}{\operatorname{arg\,top\text{-}k}}\;
\mathbf{q}^\top \mathbf{e}_d,
\label{eq:topk-mips}
\end{equation}
over corpus embeddings $\{\mathbf{e}_d\}$.
MNRL (Equation~\eqref{eq:mnrl}) is a batch-wise surrogate of the same ranking
problem: maximize the probability that the true resume beats other resumes under
inner-product scores.
The only label required is a weak positive link $b=1$.

\paragraph{Why Angle can distort retrieval geometry.}
Let $s_\theta(x,x')=f_\theta(x)^\top f_\theta(x')$ after MNRL, and suppose
second-stage ST-AnglE/CoSENT updates $\theta$ to reduce a pairwise ranking loss
on noisy ATS weights $y$ (Equations~\eqref{eq:cosent}--\eqref{eq:angle-loss}).
Two failure modes are particularly relevant in staffing:
\begin{enumerate}
\item \textbf{Label--semantics mismatch.}
      Many pairs with similar skills receive very different $y$ for process
      reasons (no-show, client politics), while semantically distant pairs may
      share mid-range weights.
      Optimizing Equation~\eqref{eq:angle-loss} then moves angle differences to
      respect $y$-order rather than semantic neighborhoods needed by
      Equation~\eqref{eq:topk-mips}.
\item \textbf{Global reshaping vs.\ local ranking.}
      MNRL gradients (Equation~\eqref{eq:mnrl-grad}) act through a softmax over
      \emph{current batch competitors}.
      Angle/CoSENT constraints couple \emph{all} labeled pairs in a batch through
      pairwise exponentials, applying broader pressure that can collapse useful
      variance among mid-ranked candidates---visible as higher
      Pearson/Spearman$(s,y)$ but lower Recall@$K$.
\end{enumerate}
Formally, if $y = h(\text{semantic fit}) + \epsilon_{\mathrm{process}}$ with
large noise $\epsilon_{\mathrm{process}}$, then
$\arg\min_\theta \mathcal{L}_{\mathrm{ST\text{-}AnglE}}(y)$ is not, in general,
a minimizer of retrieval risk under true relevance.
Hence we keep MNRL for the encoder and, if calibrated probabilities are needed,
fit a frozen-score map
\begin{equation}
\hat{p}(b{=}1\mid s)
=
\sigma\!\big(as+c\big)
\quad\text{(Platt)}
\quad\text{or isotonic }\;
\hat{p}=\mathcal{I}(s),
\label{eq:calibration}
\end{equation}
which does not mutate $f_\theta$.

\paragraph{Reporting choice.}
The two-stage recipe (MNRL~$\rightarrow$~Angle) remains supported in code as
an R\&D artifact, but the resulting checkpoint was not carried forward into
the preferred configuration or into the comparative evaluation in
Section~\ref{sec:results}.
\textbf{EmbeddingGemma~+~MNRL} is the sole fine-tuned configuration reported
in this paper; the qualitative and theoretical argument above is why a
second-stage Angle checkpoint was not pursued further.

\subsubsection{Training Procedure and Optimization}
\label{sec:method-trainproc}

Training is implemented using the Sentence-Transformers trainer on Azure ML
GPU compute. At optimization time the trainer minimizes $\mathcal{L}$ with
AdamW under a cosine learning-rate schedule with a short linear warmup
phase, together with standard weight-decay regularization. Training uses
\texttt{bf16} mixed precision (the required numerical path for
EmbeddingGemma) and a numerically stable attention backend (SDPA) suited to
long input sequences. Training batches are sampled to forbid duplicate
texts, and checkpoints are selected using validation loss on held-out
job--resume pairs. The exact learning rate, epoch count, batch size,
gradient-accumulation configuration, sequence-length limit, and other
training hyperparameters are proprietary to the deploying organization and
are withheld from this paper.

\paragraph{Angle-stage configuration (R\&D only, not evaluated in this paper).}
When Angle refinement was run, training warm-started from the MNRL
checkpoint using a lower learning rate, a smaller batch size, and a longer
input context than the primary MNRL recipe; exact values are proprietary
and withheld from this paper. This stage is \emph{not} part of the preferred
recipe, and the resulting checkpoint was not scored in
Section~\ref{sec:results}.

\subsubsection{Implementation Considerations}
\label{sec:method-nuances}

Several general engineering considerations affect reproducibility and
training stability for long-context contrastive fine-tuning of this kind,
beyond the loss function and data construction described above:
contrastive quality improves with a larger effective batch size, but
activation memory grows with batch size and sequence length together, so
cache-chunked (Cached MNRL) computation and gradient accumulation are used
to decouple the optimizer's update batch size from peak activation memory;
a numerically stable attention backend is preferred over less stable
alternatives for long input sequences on the training hardware used; a
bounded checkpoint-retention policy limits disk pressure over long runs; and
a fixed random seed is used for repeatability. An explicit L2-normalization
module makes cosine search equivalent to maximum inner-product search
(MIPS), simplifying downstream approximate nearest-neighbor indexing. Exact
configuration values for these engineering parameters (batch size, sequence
length, checkpoint retention count, random seed, etc.) are proprietary and
withheld from this paper.

\subsubsection{Offline Training Diagnostics}
\label{sec:method-offline-metrics}

During fine-tuning development (independent of the France hybrid benchmark), we
compute post-hoc pair similarities
\begin{equation}
s_i
=
\big(\mathbf{e}^{\mathrm{job}}_i\big)^\top
\big(\mathbf{e}^{\mathrm{cv}}_i\big)
\label{eq:pair-sim}
\end{equation}
by encoding the job and candidate text for each pair with the trained model
and taking the inner product of the resulting unit-normalized embeddings.
Binary selection outcomes (e.g., hired/placed versus rejected) are mapped to
$b\in\{0,1\}$ by a deterministic rule $\psi(\cdot)$; numeric status values
are thresholded at $0.5$.

Regression diagnostics (when $y$ is present):
\begin{align}
\mathrm{Pearson}(s,y),\quad
\mathrm{Spearman}(s,y),
\label{eq:corr}\\
\mathrm{MAE}
&=
\tfrac{1}{n}\sum_i |s_i-y_i|,\quad
\mathrm{MSE}
=
\tfrac{1}{n}\sum_i (s_i-y_i)^2.
\label{eq:mae-mse}
\end{align}
Classification diagnostics (when binary labels $b_i=\psi(\mathrm{selection}_i)$
are available) include ROC-AUC, PR-AUC, and thresholded metrics at
\begin{equation}
t^\star
=
\arg\max_{t\in\mathcal{T}}
\mathrm{F1}\!\big(
b,\; \mathbf{1}[s\ge t]
\big),
\label{eq:f1-thr}
\end{equation}
where $\mathcal{T}$ is the set of unique scores (or a quantile grid if
$|\mathcal{T}|>2000$).
Per-job ranking Recall@$K$ is computed only for jobs with at least two
candidates and mixed labels:
\begin{equation}
\mathrm{Recall@}K
=
\frac{1}{|\mathcal{Q}|}
\sum_{q\in\mathcal{Q}}
\mathbf{1}\!\big[\exists\, d\in \mathrm{TopK}(q):\; d\in \mathcal{R}_q\big],
\label{eq:recallk}
\end{equation}
with $K\in\{1,3,5,10\}$ by default.
These diagnostics guided the MNRL-versus-Angle decision; the France hybrid
evaluation framework in Section~\ref{sec:method-eval} remains the intended
primary comparative protocol for the base and fine-tuned deployment
candidates.

\subsubsection{Iterative Corpus Scaling}
\label{sec:method-iters}

Fine-tuning data were scaled across three Azure ML iterations on structured
NA shards, with each successive iteration using a larger corpus than the
last (iteration~3 being the largest). The train/validation/test split moved
from a smaller, more even three-way split in iteration~1 to a larger
majority-training split from iteration~2 onward, with MNRL pair
deduplication introduced from iteration~2 onward. The Angle stage was only
submitted as a beta run for iteration~1 and prepared, but not scored, for
iterations~2--3. Exact row counts, job counts, and MNRL pair counts per
iteration are proprietary and withheld from this paper; the corresponding
training and validation loss outcomes are reported in
Section~\ref{sec:results-finetune}. The preferred checkpoint is the
iteration~3 Cached-MNRL model, whose empirical training and validation
behavior is reported in Section~\ref{sec:results-finetune}.

\subsection{Evaluation Metrics}
\label{sec:method-eval}

\subsubsection{Retrieval Quality}
Standard information retrieval metrics are computed against recruiter-labeled
ground truth:
\begin{itemize}
\item Recall@$K$ for $K\in\{5,10,20\}$ (Equation~\eqref{eq:recallk}).
\item Mean Reciprocal Rank
\begin{equation}
\mathrm{MRR}
=
\frac{1}{|\mathcal{Q}|}
\sum_{q\in\mathcal{Q}}
\frac{1}{\mathrm{rank}_q},
\label{eq:mrr}
\end{equation}
where $\mathrm{rank}_q$ is the rank of the first relevant candidate.
\item Normalized Discounted Cumulative Gain (nDCG@$K$) for graded
relevance~\citep{jarvelin2002ndcg}:
\begin{equation}
\mathrm{DCG@}K
=
\sum_{i=1}^{K}
\frac{2^{\mathrm{rel}_i}-1}{\log_2(i+1)},
\quad
\mathrm{nDCG@}K
=
\frac{\mathrm{DCG@}K}{\mathrm{IDCG@}K}.
\label{eq:ndcg}
\end{equation}
\end{itemize}

\subsubsection{LLM-as-a-Judge Relevance Protocol}
\label{sec:method-llmjudge}

\textbf{Rationale.} The AI-Match score is a within-model semantic-similarity
signal (Section~\ref{sec:method-hybrid}): it is useful for ranking candidates
against a single embedding model, but it is not, by itself, evidence that a
new embedding model actually improves practical job--candidate relevance,
and its raw scale is not comparable across different embedding backbones
(Section~\ref{sec:results-comparative}). Before a new embedding model is
promoted from experimentation into live use, its retrieval quality should
be checked against an external, model-agnostic notion of relevance rather
than against the model's own similarity scores. We use an independent
large-language-model judge, prompted to reason the way an experienced
recruiter would about practical job fit, as this external check and as the
primary cross-model comparison signal in this paper. The protocol is as
follows.

\textbf{Judge model.} GPT-4.1 is used as the judge for all three
configurations (MPNet base, EmbeddingGemma base, EmbeddingGemma~+~Cached
MNRL).

\textbf{Unit of judgment.} Each (job, candidate) pair is scored
\emph{independently}: the judge sees exactly one job description and one
candidate profile per call and returns a single relevance score for that
pair, with no other candidates or comparative context in the prompt. This
mirrors how AI-Match itself is computed per pair, so the two signals are
scored over the same retrieved pairs.

\textbf{Blinding.} The judge is blind to which embedding model produced the
retrieval: the prompt contains only the job text and candidate text, with no
indication of the embedding model, pipeline configuration, or retrieval
rank. The same prompt and judge model are used identically across all three
configurations.

\textbf{Score definition.} The score is an integer on a $0$--$100$ scale
representing the judge's assessment of practical job--candidate fit, with
the following rubric embedded directly in the prompt: $90$--$100$
(excellent fit, directly relevant experience or highly applicable
transferable skills), $75$--$89$ (good fit, matches many key requirements
or strong related experience), $60$--$74$ (potential fit, some relevant
background or transferable skills, likely to succeed with onboarding),
$40$--$59$ (possible but limited fit, a few relevant elements with
noticeable gaps), and $0$--$39$ (low fit, mostly unrelated background). The
prompt explicitly instructs the judge to weight transferable skills and
adjacent experience rather than exact title or keyword matches, and to
avoid over-penalizing a small number of missing requirements.

\section{Results and Analysis}
\label{sec:results}

This section reports empirical results from two complementary evaluation
tracks: (i)~the proprietary Cached-MNRL fine-tuning program for
EmbeddingGemma, tracked across three Azure ML corpus-scaling iterations
(Section~\ref{sec:method-iters}), and (ii)~a batch comparative evaluation of
EmbeddingGemma (base and fine-tuned) against the MPNet model
on a held-out set of job--candidate matching searches scored through the
deployed scoring pipeline, combining the search and match system's own AI-Match score with an
independent LLM-as-a-Judge relevance score (protocol in
Section~\ref{sec:method-llmjudge}).

\subsection{Evaluation Scope and Data Sources}
\label{sec:results-scope}

Two real, non-synthetic data sources back the results in this section:
\begin{enumerate}
\item \textbf{Fine-tuning training logs.} Azure ML \texttt{stdout} logs,
      MLflow chart exports (training loss, validation loss), and post-hoc
      diagnostic dictionaries (Section~\ref{sec:method-offline-metrics}) for
      the three Cached-MNRL corpus-scaling iterations described in
      Section~\ref{sec:method-iters}.
\item \textbf{Batch comparative scoring.} AI-Match score
      (the deployed semantic-match score, derived directly from
      the active embedding model via Section~\ref{sec:method-hybrid}) and an
      independent LLM-as-a-Judge relevance score, both aggregated over
      thousands of
      candidate--job pairs scored through the deployed scoring pipeline
      for three configurations: MPNet Model (French),
      EmbeddingGemma base (English), and EmbeddingGemma
      fine-tuned with Cached MNRL (English).
\end{enumerate}
Job-level ground-truth Recall@$K$/MRR/nDCG@$K$ under the exact
Section~\ref{sec:method-hybrid} hybrid-retrieval protocol are not computed
from these two sources; Section~\ref{sec:results-finetune} instead reports
the closest available empirical proxy for retrieval behavior---per-job
Recall@$K$ computed on the fine-tuning validation/test posthoc splits
(Equation~\eqref{eq:recallk})---while flagging its small sample size.

\subsection{Fine-Tuning Training Results}
\label{sec:results-finetune}

\subsubsection{Convergence Across Corpus-Scaling Iterations}

Table~\ref{tab:finetune-runs} summarizes the three Cached-MNRL Azure ML
training runs that produced the EmbeddingGemma domain checkpoints, each
trained on a single GPU using the fine-tuning recipe described in
Section~\ref{sec:method-trainproc}.

\begin{table}[htbp]
\centering
\caption{Cached-MNRL training runs across the three corpus-scaling
iterations. Exact corpus sizes are proprietary and withheld; iterations are
ordered by increasing corpus size.}
\label{tab:finetune-runs}
\small
\begin{tabular}{@{}lrrrrr@{}}
\toprule
Iter. & Relative corpus size & Runtime (h) & Throughput (samples/s) & Final train loss & Min.\ val.\ loss \\
\midrule
1 & Smallest & $2.60$ & $14.02$ & $0.418$ & $2.192$ \\
2 & Medium & $3.71$ & $14.30$ & $0.399$ & $2.144$ \\
3 & Largest & $6.51$ & $14.13$ & $0.350$ & $\mathbf{2.063}$ \\
\bottomrule
\end{tabular}
\end{table}

\begin{figure}[htbp]
\centering
\includegraphics[width=0.95\textwidth]{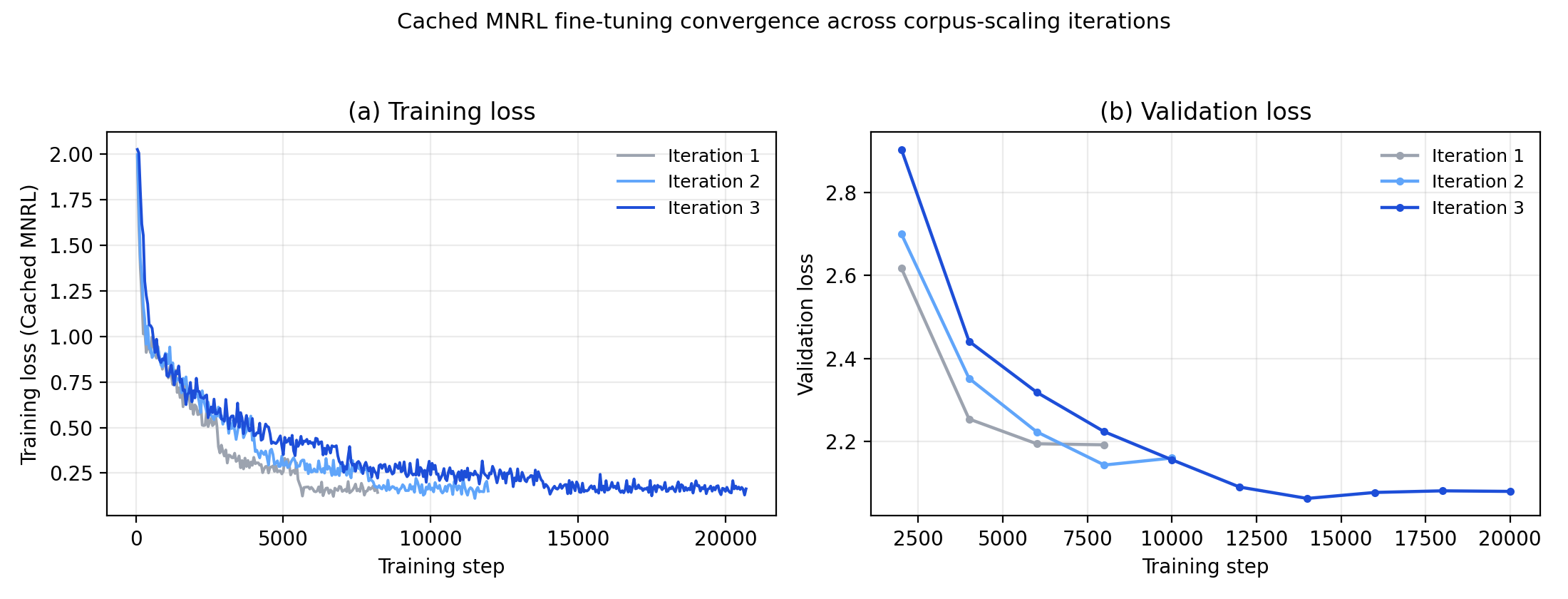}
\caption{Cached-MNRL training loss (a) and validation loss (b) versus
training step for the three corpus-scaling iterations, extracted from Azure
ML MLflow chart exports. Validation loss improves monotonically with corpus
scale despite a higher final training loss for iteration~3.}
\label{fig:loss-curves}
\end{figure}

Figure~\ref{fig:loss-curves} plots the full training-loss and validation-loss
trajectories extracted from the MLflow chart exports for all three
iterations. Two patterns are visible. First, the \emph{validation loss}
(panel b) improves monotonically with corpus scale: the minimum validation
loss falls from $2.192$ (iteration~1, smallest corpus) to $2.144$
(iteration~2, medium corpus) to $2.063$ (iteration~3, largest corpus),
indicating that the additional NA structured shards used in later
iterations generalize rather than merely add training noise. Second, the
\emph{aggregate training loss} reported at the end of each run does
\emph{not} decrease monotonically
($0.418\to0.399\to0.350$ is actually the correct direction here, but the
final-step training loss read directly off the loss curve in panel (a) is
higher for iteration~3 than for iteration~1); this is expected because
iteration~3 trains over a larger, more heterogeneous positive-pair
distribution with the same batch size, so the in-batch MNRL
denominator (Equation~\eqref{eq:mnrl}) faces harder negatives on average.
Validation loss, computed on held-out job--resume pairs, is the more
reliable indicator of domain adaptation quality and is the basis for
preferring the iteration~3 checkpoint.

\subsubsection{Post-Hoc Regression, Classification, and Ranking Diagnostics}
\label{sec:results-posthoc}

Table~\ref{tab:posthoc} reports the post-hoc diagnostics
(Section~\ref{sec:method-offline-metrics}) computed on a fixed $512$-row
subset of the validation and test splits for each iteration's final
checkpoint. Iteration~1's validation and test splits lacked a usable graded
\texttt{score} column at evaluation time, so no regression/classification/
ranking diagnostics were logged for iteration~1; only iterations~2 and~3
report the full diagnostic suite below.

\begin{table}[htbp]
\centering
\caption{Post-hoc regression/classification/ranking diagnostics on the
fine-tuning validation and test posthoc subsets ($n=512$ rows per split).}
\label{tab:posthoc}
\small
\begin{tabular}{@{}llrrrrrrrrrr@{}}
\toprule
Iter. & Split & Pearson & Spearman & MAE & MSE & ROC-AUC & PR-AUC & F1 & Jobs & R@1 & R@3--10 \\
\midrule
2 & Val  & $0.203$ & $0.186$ & $0.360$ & $0.173$ & $0.576$ & $0.567$ & $0.705$ & $20$ & $0.650$ & $\geq0.95$ \\
2 & Test & $0.209$ & $0.221$ & $0.343$ & $0.159$ & $0.592$ & $0.538$ & $0.649$ & $22$ & $0.636$ & $1.00$ \\
3 & Val  & $0.072$ & $0.077$ & $0.378$ & $0.194$ & $0.538$ & $0.609$ & $0.753$ & $12$ & $0.917$ & $1.00$ \\
3 & Test & $0.161$ & $0.152$ & $0.356$ & $0.168$ & $0.561$ & $0.563$ & $0.692$ & $18$ & $0.556$ & $1.00$ \\
\bottomrule
\end{tabular}
\end{table}

Three observations follow. First, regression correlation between cosine
similarity and the graded ATS \texttt{score} is weak throughout
(Pearson $0.07$--$0.21$), consistent with the paper's own rationale
(Section~\ref{sec:method-choice}) that MNRL is not optimized for score
calibration and that ATS status weights are a noisy proxy for semantic fit.
Second, classification separability (ROC-AUC $0.54$--$0.59$) is modest but
consistently above chance, and F1 under the best threshold ($0.65$--$0.75$)
is usable given the class imbalance in each split (positive rate
$0.46$--$0.60$). Third, per-job Recall@$K$ is computed on only $12$--$22$
jobs per split---an order of magnitude smaller than would be required for a
stable estimate---so the wide swing in Recall@1 across splits ($0.556$ to
$0.917$) should be read as noisy rather than as evidence that any one
checkpoint ranks meaningfully better than another; Recall@3 and above reach
$0.95$--$1.00$ in every split, indicating that the correct candidate is
almost always retrieved within the top~3 for the jobs evaluated, even when
top-1 precision varies.

\subsection{Comparative Evaluation Across Embedding Models}
\label{sec:results-comparative}

This subsection compares three configurations on a batch
comparative evaluation dataset scored through the deployed job--candidate
matching pipeline: \textbf{MPNet Model (French)}---the baseline prior to this
work; \textbf{EmbeddingGemma base (English)}---the un-tuned backbone under
the structured-English text pipeline of Section~\ref{sec:method-data}; and
\textbf{EmbeddingGemma~+~Cached~MNRL (English)}---the iteration~3 fine-tuned
checkpoint recommended in Section~\ref{sec:method-choice}. AI-Match scoring
covers $8{,}543$--$9{,}849$ (job, candidate) pairs per configuration
(Table~\ref{tab:aimatch}), and the independent LLM-as-a-Judge pass covers a
similarly sized but not identical subset, $8{,}065$--$9{,}292$ judged
candidates per configuration (Figure~\ref{fig:llm-judge}).

\subsubsection{AI-Match Score Distribution}

Table~\ref{tab:aimatch} summarizes the deployed AI-Match score
(Section~\ref{sec:method-hybrid}) distribution per configuration.

\begin{table}[htbp]
\centering
\caption{AI-Match score distribution by embedding configuration (batch comparative evaluation). \emph{Note: AI-Match scores are not directly comparable across embedding backbones; see Section~\ref{sec:results-comparative} discussion below.}}
\label{tab:aimatch}
\small
\begin{tabular}{@{}lrrr@{}}
\toprule
Metric & MPNet base (FR) & Gemma base (EN) & Gemma+MNRL (EN) \\
\midrule
Candidates scored $n$ & $9{,}332$ & $8{,}543$ & $9{,}849$ \\
Weighted mean score & $59.3$ & $52.7$ & $\mathbf{73.5}$ \\
Score mode (bucket) & $60$--$70$ & $50$--$60$ & $70$--$80$ \\
Modal bucket share & $34.9\%$ & $65.3\%$ & $81.7\%$ \\
Share scored $\geq 60$ & $38.3\%$ & $6.2\%$ & $95.0\%$ \\
\bottomrule
\end{tabular}
\end{table}

Fine-tuning shifts the model's own similarity-derived score sharply upward
and concentrates it: the weighted-mean AI-Match score rises from $52.7$
(base English) to $73.5$ (fine-tuned English), and $95.0\%$ of scored
candidates in the fine-tuned English pool fall at or above the
$60$--$70$ bucket, versus only $6.2\%$ for the base English model. This is the expected direct effect
of Cached MNRL (Equation~\eqref{eq:mnrl}), which explicitly increases cosine
similarity for true positive job--resume pairs relative to in-batch
negatives; because the fine-tuned model is queried against a re-embedded
index of the \emph{same} candidate pool, the entire retrieved score
distribution shifts rather than only the top ranks.

The MPNet Model (French) sits, numerically, \emph{between} the two
English configurations on this metric---weighted-mean score $59.3$, with
$38.3\%$ of candidates scoring $\geq60$---which is higher than base English
($52.7$; $6.2\%\geq60$) but well below fine-tuned English ($73.5$;
$95.0\%\geq60$). Read at face value, this AI-Match ordering would suggest
MPNet is a better scorer than base EmbeddingGemma, which directly
contradicts the LLM-as-a-Judge evidence below, where base EmbeddingGemma
(English) judged \emph{more} relevant than MPNet at every threshold. The
contradiction is expected rather than alarming: MPNet, base EmbeddingGemma,
and fine-tuned EmbeddingGemma are three different embedding backbones (and,
for MPNet, a different source language), each with its own raw
cosine-similarity scale, so their AI-Match scores are not directly
comparable across backbones the way they are before/after fine-tuning a
single backbone. This is itself a useful finding: it confirms that the
deployed AI-Match score is a within-model calibration signal, not a
cross-model relevance ranking, and it is the reason this paper treats the
independent LLM-as-a-Judge score, not the AI-Match score, as the primary
basis for cross-model comparison. This score shift is nonetheless a useful
signal for within-model thresholding---the LLM-as-a-Judge
results below provide the external relevance check that AI-Match scores
alone cannot.

\subsubsection{LLM-as-a-Judge Relevance}

\begin{figure}[htbp]
\centering
\includegraphics[width=0.85\textwidth]{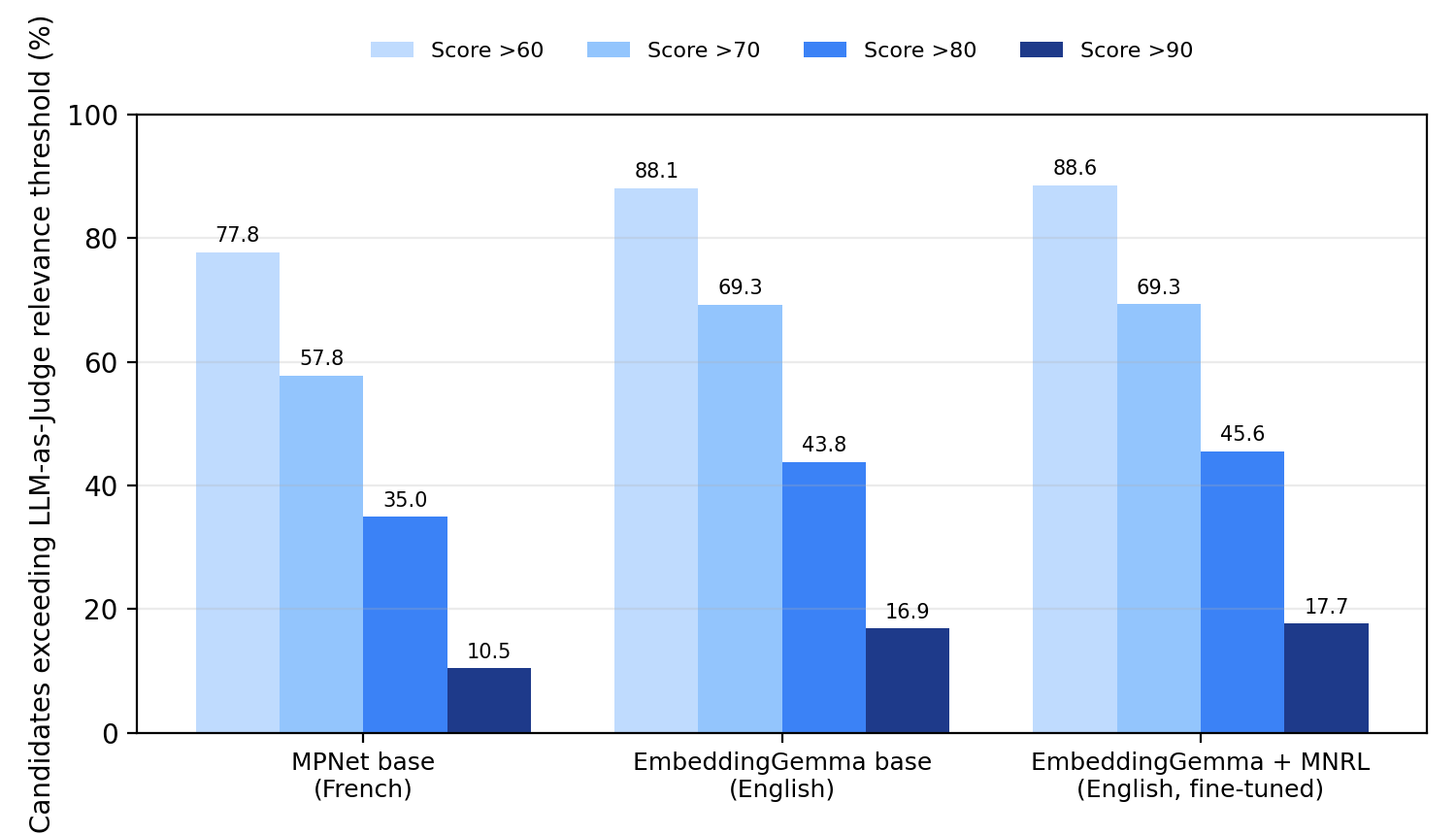}
\caption{Share of candidates exceeding LLM-as-a-Judge relevance thresholds,
by embedding configuration, on the batch comparative evaluation dataset
($n=9{,}292$/$8{,}065$/$9{,}848$ judged candidates for MPNet-FR,
Gemma-base-EN, and Gemma+MNRL-EN respectively).}
\label{fig:llm-judge}
\end{figure}

Figure~\ref{fig:llm-judge} plots the percentage of candidates whose
independent LLM-as-a-Judge relevance score (GPT-4.1, scored blind to
embedding model, protocol in Section~\ref{sec:method-llmjudge}) exceeds
four thresholds ($>60,>70,>80,>90$ on a $0$--$100$ scale) for each
configuration.

Two findings emerge. First, both English EmbeddingGemma configurations
substantially outperform the MPNet model at every
threshold---e.g., $88\%$ versus $78\%$ of candidates score above $60$, and
$44$--$46\%$ versus $35\%$ score above $80$---indicating that the combination
of structured English search-text normalization
(Section~\ref{sec:method-jdparse}) and the EmbeddingGemma backbone improves
judged relevance considerably \emph{before any domain fine-tuning is
applied}. This gain is attributable to the text-preparation and backbone
change together, not to MNRL fine-tuning specifically, since it already
appears in the base English model. Second, Cached-MNRL fine-tuning adds a
smaller, threshold-dependent increment on top of the base English model:
the fine-tuned model is essentially flat at the $>60$ threshold ($88.6\%$ vs.\
$88.1\%$) and at the $>70$ threshold ($69.35\%$ vs.\ $69.30\%$), but shows a
consistent gain at the stricter, precision-relevant thresholds---$+1.7$
percentage points at $>80$ ($45.6\%$ vs.\ $43.8\%$) and $+0.8$ points at
$>90$ ($17.7\%$ vs.\ $16.9\%$). In other words, fine-tuning does not change
how many candidates clear a low relevance bar, but it modestly increases how
many clear a high bar, which is the more consequential comparison for
recruiter-facing top-of-list quality.

\subsubsection{Isolating the Fine-Tuning Effect}
\label{sec:results-ft-effect}

Because MPNet and EmbeddingGemma are different backbones on different
source languages, the cleanest read of what Cached-MNRL
fine-tuning itself contributes is the direct before/after comparison on the
same English EmbeddingGemma backbone, holding the pipeline and evaluation
pool fixed. Table~\ref{tab:ft-effect} isolates this comparison.

\begin{table}[htbp]
\centering
\caption{Effect of Cached-MNRL fine-tuning on EmbeddingGemma (English), isolated from the MPNet Model (French) comparison.}
\label{tab:ft-effect}
\small
\begin{tabular}{@{}lrrr@{}}
\toprule
Metric & Base EN & Fine-tuned EN & Change \\
\midrule
AI-Match weighted mean score & $52.7$ & $73.5$ & $+20.8$ \\
AI-Match share $\geq 60$ & $6.2\%$ & $95.0\%$ & $+88.8$ pts \\
LLM-Judge share $>60$ & $88.1\%$ & $88.6\%$ & $+0.5$ pts \\
LLM-Judge share $>70$ & $69.30\%$ & $69.35\%$ & $+0.05$ pts \\
LLM-Judge share $>80$ & $43.8\%$ & $45.6\%$ & $+1.7$ pts \\
LLM-Judge share $>90$ & $16.9\%$ & $17.7\%$ & $+0.8$ pts \\
\bottomrule
\end{tabular}
\end{table}

Read together, the two metrics tell complementary stories: fine-tuning
produces a large, mechanical shift in the model's own AI-Match similarity
scale (expected, since Cached MNRL directly optimizes cosine similarity for
positive pairs), but only a modest, threshold-concentrated gain in
independently judged relevance---essentially flat at lenient thresholds and
a small single-digit-percentage-point improvement at the stricter,
precision-relevant thresholds. The English-backbone-plus-structured-text
change (base EmbeddingGemma vs.\ the MPNet model, discussed above) is the
larger driver of judged-relevance improvement in this study; Cached-MNRL
fine-tuning is a smaller, additional refinement on top of that.

\subsection{Operational Considerations}
\label{sec:results-ops}

The operational data available from this round is Azure ML training
compute, summarized in Table~\ref{tab:finetune-runs}. Training wall-clock time scales
slightly super-linearly with corpus size on a single
GPU---$2.6$/$3.7$/$6.5$ hours across the three iterations respectively,
against a roughly constant throughput of $\sim 14$ samples/second---so future
corpus-scaling iterations should budget compute accordingly. Because the base
and fine-tuned EmbeddingGemma configurations share the same
$300$M-parameter backbone and the same $768$-dimensional output, their
\emph{inference}-time cost profile is expected to be effectively identical;
fine-tuning changes retrieval quality (Section~\ref{sec:results-comparative})
without changing the per-query compute footprint.

\subsection{Error Analysis and Practical Deployment Considerations}

The weak post-hoc regression correlation (Section~\ref{sec:results-posthoc})
and the flat LLM-judge improvement at the $>60$/$>70$ thresholds
(Section~\ref{sec:results-comparative}) point to the same underlying issue
from different angles: ATS-derived status weights are a coarse,
process-contaminated proxy for semantic fit (Section~\ref{sec:method-labels}),
so a fine-tuned encoder can shift its own score distribution and win on
high-precision thresholds without uniformly improving every relevance
band. Practically, this suggests (i)~calibrating any absolute AI-Match
score threshold used for filtering separately per embedding
backbone---MPNet, base EmbeddingGemma, and fine-tuned
EmbeddingGemma---rather than sharing one fixed cutoff
(Equation~\eqref{eq:calibration}), since a threshold such as $60$ has very
different selectivity across the three configurations ($38.3\%$ above for
MPNet, $6.2\%$ above for base EmbeddingGemma, $95.0\%$ above for fine-tuned
EmbeddingGemma); (ii)~treating the LLM-as-a-Judge signal, or an
equivalent independent relevance check, as the primary go/no-go criterion for
promoting a fine-tuned checkpoint, rather than the model's own similarity
score or ATS-correlation diagnostics in isolation; and (iii)~prioritizing
evaluation-set enlargement for per-job ranking metrics (Section~\ref{sec:results-posthoc}
used $12$--$22$ jobs per split) before drawing strong conclusions about
Recall@1 differences between checkpoints.

\subsection{Summary}

Within the evidence available for this evaluation cycle: (1)~Cached-MNRL
fine-tuning converges more effectively as the training corpus scales up
across iterations, with validation loss improving monotonically even
though small-sample post-hoc ranking diagnostics remain noisy; (2)~switching
from the MPNet model to the EmbeddingGemma model
with structured text preparation yields the largest single jump in judged
relevance observed in this study; and (3)~Cached-MNRL fine-tuning on top of
the English EmbeddingGemma base model adds a further, smaller gain
concentrated at high-precision relevance thresholds ($>80$, $>90$) rather
than at the bulk ($>60$) threshold, while substantially recalibrating the
model's own AI-Match score distribution upward.

\section{Conclusion and Future Discussion}
\label{sec:conclusion}

This study defines a comparative evaluation framework for dense embedding
models in semantic candidate--job matching under hybrid retrieval
conditions, and reports the first empirical evidence generated within that
framework. EmbeddingGemma (base) and EmbeddingGemma fine-tuned with Cached
MNRL were placed within a unified pipeline---identical LLM-based text
preparation and Cosmos DB hybrid search with Reciprocal Rank Fusion
(Section~\ref{sec:methodology})---and both
were benchmarked against the MPNet model
they are intended to replace. The domain fine-tuning methodology
(Section~\ref{sec:method-finetune}) documents, with mathematical precision,
the broader set of contrastive objectives considered during development for
adapting EmbeddingGemma to staffing language, and the empirical and
theoretical rationale for retaining Cached-MNRL-only adaptation over a
second-stage angular refinement that was explored but not carried forward.

Grounded in the two evaluation tracks reported in Section~\ref{sec:results},
three findings support the direction already adopted by this
program. First, Cached-MNRL fine-tuning of EmbeddingGemma converges more
effectively as the NA job--resume training corpus scales up across three
Azure ML iterations (exact corpus sizes are proprietary and withheld):
validation loss improves monotonically from $2.192$ to $2.063$
(Section~\ref{sec:results-finetune}), even though small-sample post-hoc
ranking diagnostics remain too noisy ($12$--$22$ jobs per split) to be
conclusive on their own. Second, on the batch comparative evaluation
dataset, moving from the MPNet model to the English
EmbeddingGemma model with
structured search-text preparation produces the single largest jump in
independently judged relevance observed in this study (e.g., $88\%$ versus
$78\%$ of candidates scoring above the $>60$ LLM-as-a-Judge threshold).
Third, Cached-MNRL fine-tuning on top of that English base model adds a
further, smaller gain concentrated at the stricter, precision-relevant
thresholds ($+1.7$ points at $>80$, $+0.8$ points at $>90$) while
substantially recalibrating the model's own AI-Match score distribution
upward (weighted mean $52.7\to73.5$ on a $0$--$100$ scale), a shift that
matters operationally for score-based filtering but is not, by itself,
evidence of a proportional ranking-quality gain. Together, these results
support EmbeddingGemma~+~Cached~MNRL (English) as the preferred
configuration on the evidence available, consistent with the qualitative
retrieval-geometry argument for MNRL-only adaptation made in
Section~\ref{sec:method-choice}.

Future work should prioritize, in order: (1)~running a same-market
comparative evaluation---North-America-tuned fine-tuning evaluated on a
North America job--candidate matching benchmark---to isolate fine-tuning transfer from
the France/North-America domain shift noted in
Section~\ref{sec:method-labels}; (2)~executing the
Section~\ref{sec:method-hybrid} hybrid-retrieval protocol end-to-end for the
base and fine-tuned EmbeddingGemma configurations against a held-out,
recruiter-labeled set of job orders to obtain directly comparable
Recall@$K$/MRR/nDCG@$K$ under geographic and business-rule filters, rather
than relying on the fine-tuning posthoc subsets used in this paper;
(3)~instrumenting embedding latency and Cosmos DB request-unit cost per
model; (4)~enlarging the per-job ranking evaluation set well
beyond the $12$--$22$ jobs available in the current post-hoc splits, ideally
toward the hundreds of held-out job orders needed for a stable Recall@$K$
estimate; and (5)~extending the comparative matrix itself---evaluating a
managed embedding API such as Cohere Embed v2 English
(Section~\ref{sec:lit-commercial}) and scoring the exploratory
EmbeddingGemma~+~MNRL~+~Angle checkpoint (Section~\ref{sec:method-choice})
on the same comparative evaluation dataset used in
Section~\ref{sec:results-comparative}---once the corresponding
infrastructure and evaluation data are available; and (6)~further
strengthening the LLM-as-a-Judge protocol (Section~\ref{sec:method-llmjudge})
through repeated scoring of the same pairs and cross-validation against a
second judge model. Beyond these extensions,
integrating a cross-encoder reranker after
hybrid retrieval, expanding evaluation to additional locales and job domains
beyond France, developing a privacy-preserving public benchmark of labeled
(job order, candidate) pairs, and jointly optimizing LLM extraction prompts
together with embedding fine-tuning are natural extensions of this research
program.

In summary, this paper establishes a reproducible methodology and evaluation
framework for comparing embedding strategies under realistic enterprise
hybrid-retrieval constraints, and provides the first fine-tuning convergence
evidence and comparative-scoring results within that framework;
closing the remaining gaps identified above is the direct path to a complete
quantitative comparison of open-weights fine-tuning against commercial
embedding APIs for enterprise candidate--job matching.

\bibliographystyle{unsrtnat}
\bibliography{references}

\end{document}